\documentclass[runningheads]{llncs}
\AtBeginDocument{%
  \providecommand\BibTeX{{%
    \normalfont B\kern-0.5em{\scshape i\kern-0.25em b}\kern-0.8em\TeX}}}

\usepackage{graphicx}
\usepackage[labelfont=bf]{caption}
\usepackage{subcaption}
\usepackage{booktabs}
\usepackage{listings}
\usepackage{color}
\usepackage{xcolor}
\usepackage{pythonhighlight}
\definecolor{codegreen}{rgb}{0,0.6,0}
\definecolor{codegray}{rgb}{0.5,0.5,0.5}
\definecolor{codepurple}{rgb}{0.58,0,0.82}
\definecolor{backcolour}{rgb}{0.95,0.95,0.92}

\lstdefinestyle{mystyle}{
    backgroundcolor=\color{backcolour},   
    commentstyle=\color{codegreen},
    keywordstyle=\color{magenta},
    numberstyle=\tiny\color{codegray},
    stringstyle=\color{codepurple},
    basicstyle=\ttfamily\footnotesize,
    breakatwhitespace=false,         
    breaklines=true,                 
    captionpos=b,                    
    keepspaces=true,                 
    numbers=left,                    
    numbersep=5pt,                  
    showspaces=false,                
    showstringspaces=false,
    showtabs=false,                  
    tabsize=2
}

\titlerunning{Container Orchestration in Edge and Fog Computing Environments}
\begin{document}

\title{Container Orchestration in Edge and Fog Computing Environments for Real-Time IoT Applications}

\author{Zhiyu Wang, Mohammad Goudarzi, Jagannath Aryal, and Rajkumar Buyya}
\institute{Cloud Computing and Distributed Systems (CLOUDS) Laboratory,\\
  School of Computing and Information Systems,\\ The University of Melbourne, Australia}

\maketitle



\begin{abstract}
Resource management is the principal factor to fully utilize the potential of Edge/Fog computing to execute real-time and critical IoT applications. Although some resource management frameworks exist, the majority are not designed based on distributed containerized components. Hence, they are not suitable for highly distributed and heterogeneous computing environments. Containerized resource management frameworks such as FogBus2 enable efficient distribution of framework's components alongside IoT applications' components. However, the management, deployment, health-check, and scalability of a large number of containers are challenging issues. To orchestrate a multitude of containers, several orchestration tools are developed. But, many of these orchestration tools are heavy-weight and have a high overhead, especially for resource-limited Edge/Fog nodes. Thus, for hybrid computing environments, consisting of heterogeneous Edge/Fog and/or Cloud nodes, lightweight container orchestration tools are required to support both resource-limited resources at the Edge/Fog and resource-rich resources at the Cloud. Thus, in this paper, we propose a feasible approach to build a hybrid and lightweight cluster based on K3s, for the FogBus2 framework that offers containerized resource management framework. This work addresses the challenge of creating lightweight computing clusters in hybrid computing environments. It also proposes three design patterns for the deployment of the FogBus2 framework in hybrid environments, including 1) Host Network, 2) Proxy Server, and 3) Environment Variable. The performance evaluation shows that the proposed approach improves the response time of real-time IoT applications up to 29\% with acceptable and low overhead.
\end{abstract}


\keywords{Edge Computing, Fog Computing, Container Orchestration, 
Internet of Things, Resource Management Framework}



\section{Introduction}
With the rapid development of hardware, software, and communication technology, IoT devices have become dominant in all aspects of our lives. Traditional physical devices are connected in the Internet of Things (IoT) environment to perform humanoid information perception and collaborative interaction. They realize self-learning, processing, decision-making, and control, thereby completing intelligent production and service and promoting the innovation of people's life and work patterns \cite{gubbi2013internet}. 
\par
On this premise, Cloud computing, with its powerful computing and storage capabilities, becomes a shared platform for IoT big data analysis and processing. In most cases, IoT devices offload complex applications to the Cloud for storage and processing, and the output results are then sent from the Cloud to IoT devices \cite{goudarzi2021distributed,aazam2014cloud}. As a result, users do not have concerns about insufficient storage space or computational capacity for IoT devices. However, with the explosive growth in the number of IoT devices, nowadays, the amount of raw data sensed and acquired by the IoT has been significantly increasing. Consequently, filtering, processing, and analyzing the massive amount of data in a centralized approach has become an inevitable challenge for the cloud computing paradigm \cite{aazam2014cloud,goudarzi2020application}.
\par
Moreover, the number of real-time IoT applications has been significantly increased. These applications require resources that support fast processing and low access latency to minimize the total response time \cite{goudarzi2021distributedMigration}. Some examples of these applications are autonomous robots and disaster management applications (e.g., natural hazard management).
    
\begin{figure}[!t]
  \centering
  \includegraphics[width=0.7\linewidth]{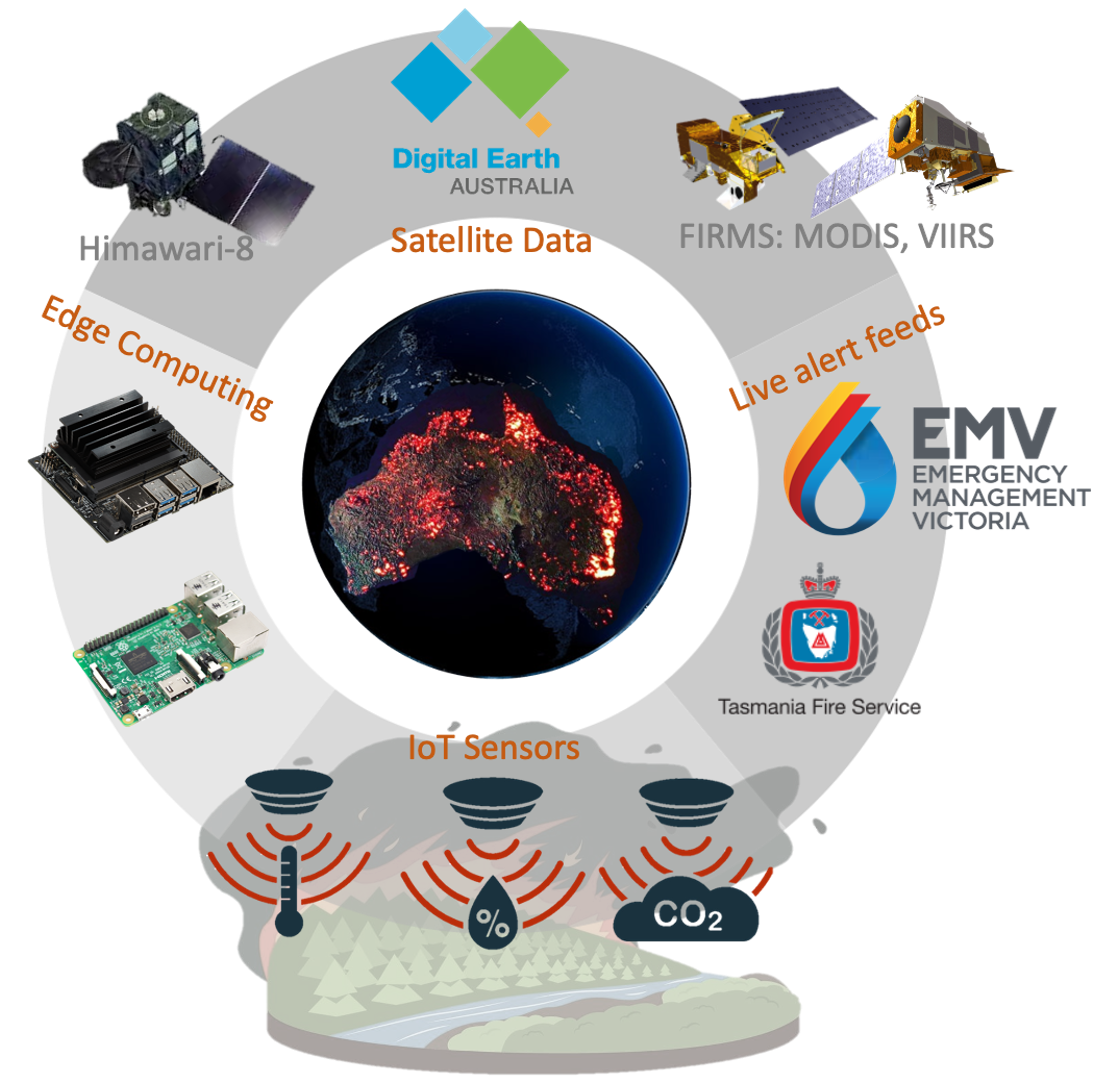}
  \caption{A visualisation framework on how satellite and ground-based sensors can be fused utilising distributed computing paradigm such as edge computing in providing accurate real-time information to end users.}
  \label{Fig:NDM_1}
\end{figure}
\begin{figure*}[!t]
  \centering
  \includegraphics[width=\linewidth]{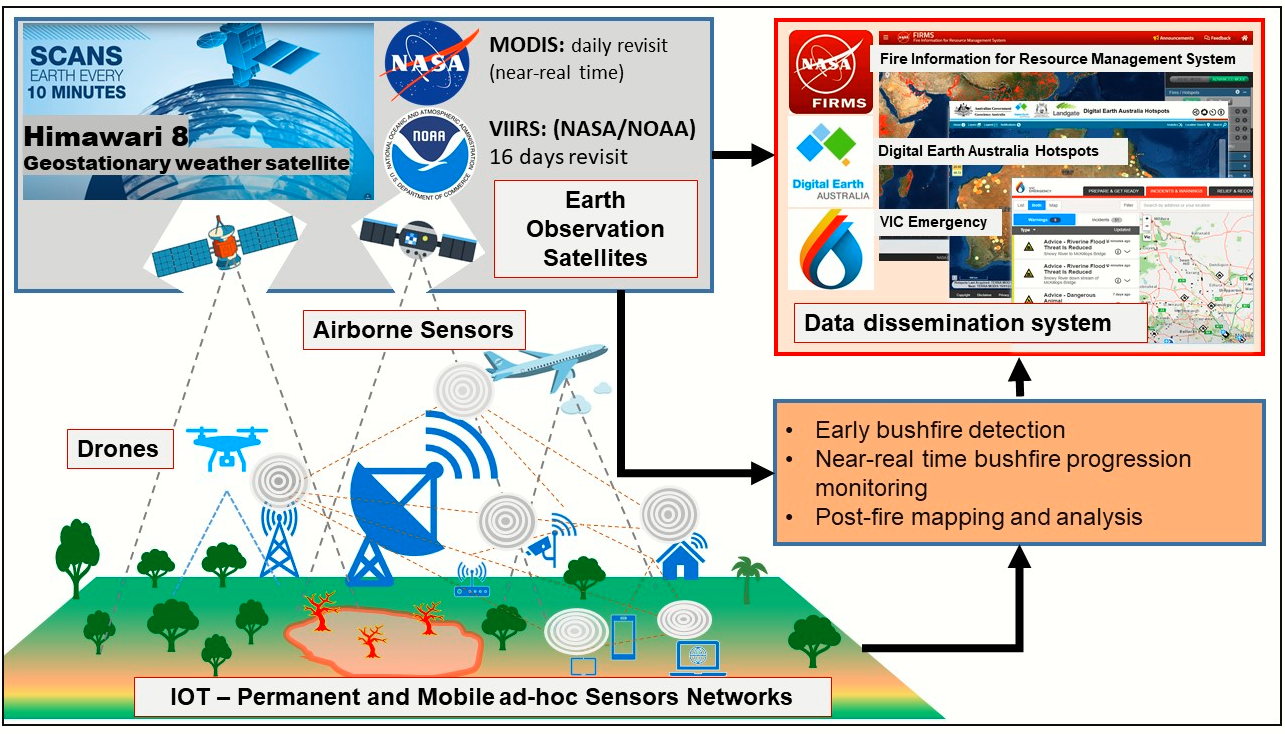}
  \caption{A detailed system model on various sensors integration and their utilisation in disseminating data to inform end users.}
  \label{Fig:NDM_2}
\end{figure*}
\subsection{Case Study: Natural Disaster Management (NDM)}
NDM comprises four phases, namely Prevention, Preparedness, Response, and Recovery. It is commonly referred to as the PPRR framework for disaster management. These four phases are not linear and independent as they overlap and support each other for a better balance between risk reduction and community resilience for better response and effective recovery. Geo-spatial solutions for different phases are in offer in practice considering the availability of big earth observation satellite data achieved from various satellite missions and IoT enabled ground-based sensor information \cite{ujjwal2019cloud}. However, the optimal fusion of satellite-based sensors and IoT sensors can provide accurate and precise information in the case of natural disasters. 
As presented in Figure~\ref{Fig:NDM_1} and Figure~\ref{Fig:NDM_2}, for the case of bushfire problems in Australia, satellite-based sensors and IoT-based sensors have been used in an ad-hoc manner to inform the end-users. For example, repositories of satellite data primarily from NASA and Digital Earth Australia and other location-based data are being used for the live alert feeds by the emergency services in different states. The potential of satellite data and their fusion in extracting the optimal information in real-time is a challenge due to the granularity of sensor-specific spatial data structure on spatial, spectral, temporal, and radiometric resolutions. With the IoT-based real-time information, there is a strong potential to validate and calibrate the satellite information captured in different resolutions to inform bush-fire alerts in space and time. For example, early bushfire detection, near-real-time bushfire progression monitoring, and post-fire mapping and analysis are possible with the optimal integration of ground-based sensors to the satellite-based sensor's information. The framework of integrating sensors and providing accurate information to end-users in real-time will help in saving lives and properties. 
\subsection{Edge and Fog Computing}
\begin{figure}[ht]
  \centering
  \includegraphics[width=\linewidth]{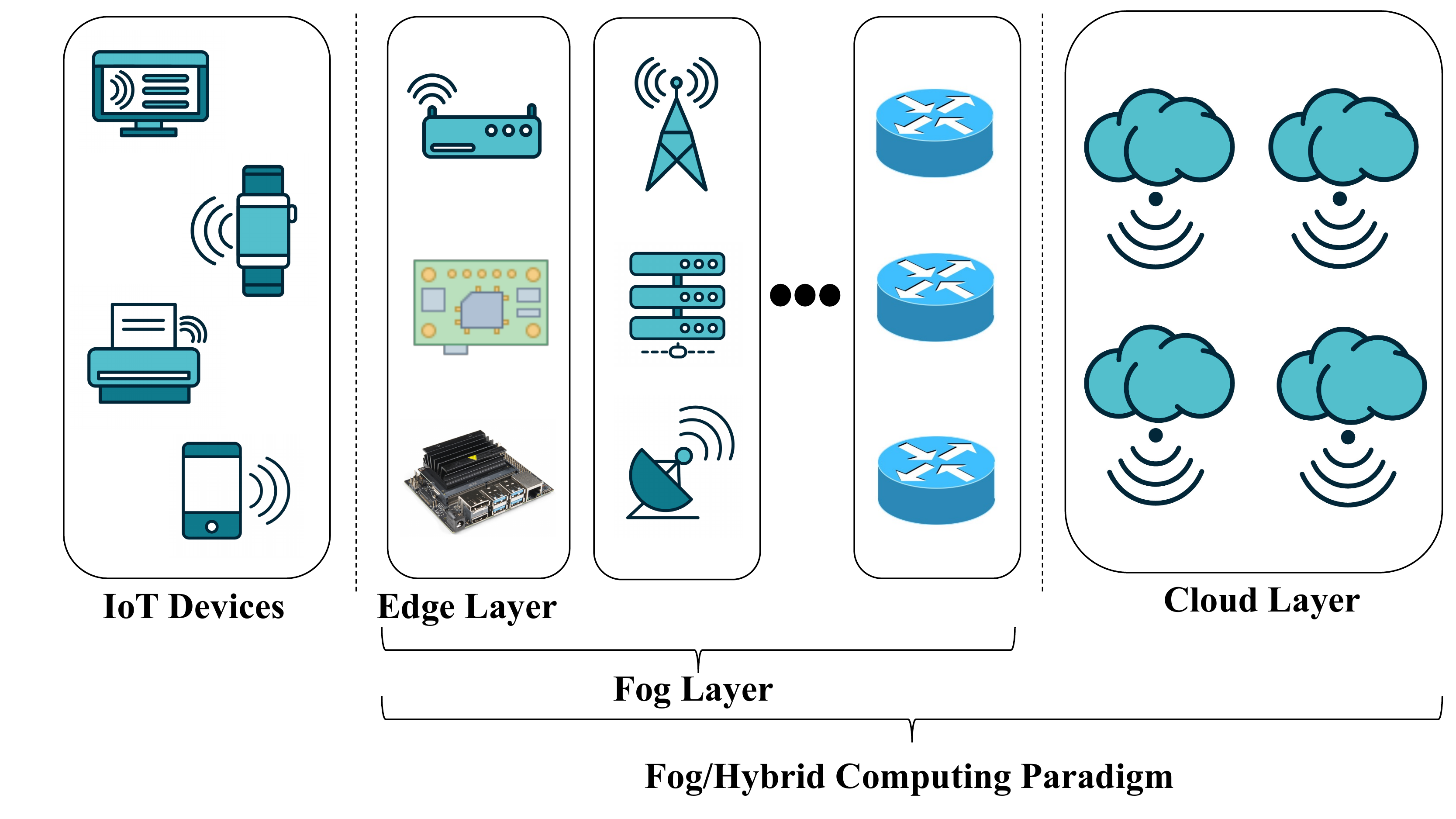}
  \caption{Overview of Fog/Hybrid computing environment}
  \label{Fig:Fogcom}
\end{figure}

For smooth and efficient execution of IoT applications, distributed computing paradigms, called Edge and Fog computing, have been emerged. They concentrate data and processing units as close as possible to the end-users, as opposed to the traditional cloud computing paradigm that concentrates data and processing units in cloud data centers \cite{buyya2019fog}. The key idea behind Edge and Fog computing is to bring Cloud-like services to the edge of the network, resulting in less application latency and a better quality of experience for users \cite{dastjerdi2016fog,goudarzi2019fog}. Edge computing can cope with medium to lightweight tasks. However, when the users' requirements consist of complex and resource-hungry tasks, Edge devices are often unable to efficiently satisfy those requirements since they have limited computing resources \cite{buyya2019fog,shi2016edge}. To address these challenges, Fog computing, also referred to as hybrid computing, is becoming a popular solution. 
Figure~\ref{Fig:Fogcom} depicts an overview of the Fog/Hybrid computing environment. In our view, Edge computing only harnesses the closest resources to the end-users while Fog computing uses deployed resources at Edge and Cloud layers. In such computing environments, Cloud can act as an orchestrator, which is responsible for big and long-period data analysis. It can operate in areas such as management, cyclical maintenance, and execution of computation-intensive tasks. Fog computing, on the other hand, efficiently manages the analysis of real-time data to better support the timely processing and execution of latency-sensitive tasks. However, in practice, contradicting the strong market demand, Fog computing is still in its infancy, with problems including no unified architecture, the large number and wide distribution of Edge/Fog nodes, and lack of technical standards and specifications. 
\par
Meanwhile, container technology has been significantly developing in recent years. Compared with physical and virtual machines, containers are very lightweight, simple to deploy, support multiple architectures, have a short start-up time, and are easy to expand and migrate. These features provide a suitable solution to the problem of severe heterogeneity of Edge/Fog nodes \cite{bali2019rule}. Container technology is being dominantly used by industry and academia to run commercial, scientific, and big data applications, build IoT systems and deploy distributed containerized resource management frameworks such as FogBus2 framework \cite{deng2021fogbus2}. FogBus2, which is a distributed and containerized framework, enables fast and efficient resource management in hybrid computing environments.
\par
Considering the ever-increasing number of containerized applications and frameworks, efficient management and orchestration of resources have become an important challenge. While container orchestration tools such as Kubernetes have become the ideal solution for managing and scaling deployments, nodes, and clusters in the industry today \cite{cai2021inverse}, there are still many challenges with their practical deployments in hybrid computing environments. Firstly, orchestration techniques need to consider the heterogeneity of computing resources in different environments for complete adaptability. Secondly, the complexity of installing and configuring hybrid computing environments should be addressed when implementing orchestration techniques. Thirdly, a strategy needs to be investigated to solve potential conflicts between orchestration techniques and the network model in the hybrid computing environment. Also, as Edge/Fog devices are resource-limited, lightweight orchestration techniques should be deployed to free up the resources for the smooth execution of end-user applications. Finally, integrating containerized resource management frameworks with lightweight orchestration tools is another important yet challenging issue to support the execution of a diverse range of IoT applications.
\par
To address these problems, this paper investigates the feasibility of deploying container orchestration tools in hybrid computing environments to enable scalability, health checks, and fault tolerance for containers.
\par
The main contributions of this paper can be summarized as follows:
\begin{itemize}
\item Presents feasible designs for implementing container orchestration techniques in hybrid computing environments. 
\item Proposes three design patterns for the deployment of the FogBus2 framework using container orchestration techniques.
\item Puts forward the detailed configurations for the practical deployment of the FogBus2 framework using container orchestration tools.
\end{itemize}
\par
The rest part of the paper is organized as follows. Section~\ref{Sec:related_works} provides a background study on the relevant technologies and reviews the container orchestration techniques in Fog computing environments. Section~\ref{Sec:Orchestration} describes the configuration properties of the K3s cluster and the detailed implementation of deploying the FogBus2 framework into the K3s cluster. Section~\ref{Sec:performance_evaluation} presents the performance evaluation. Finally, Section~\ref{Sec:conclusion} concludes the paper and presents future directions.
\section{Background Technologies and Related Work}
\label{Sec:related_works}
This section discusses the resource management framework and container orchestration tools, including the FogBus2 framework and K3s. Moreover, it also reviews the existing works on container orchestration in the cloud and Edge/Fog computing environments.
\subsection{FogBus2 Framework}
\begin{figure}[ht]
  \centering
  \includegraphics[width=\linewidth]{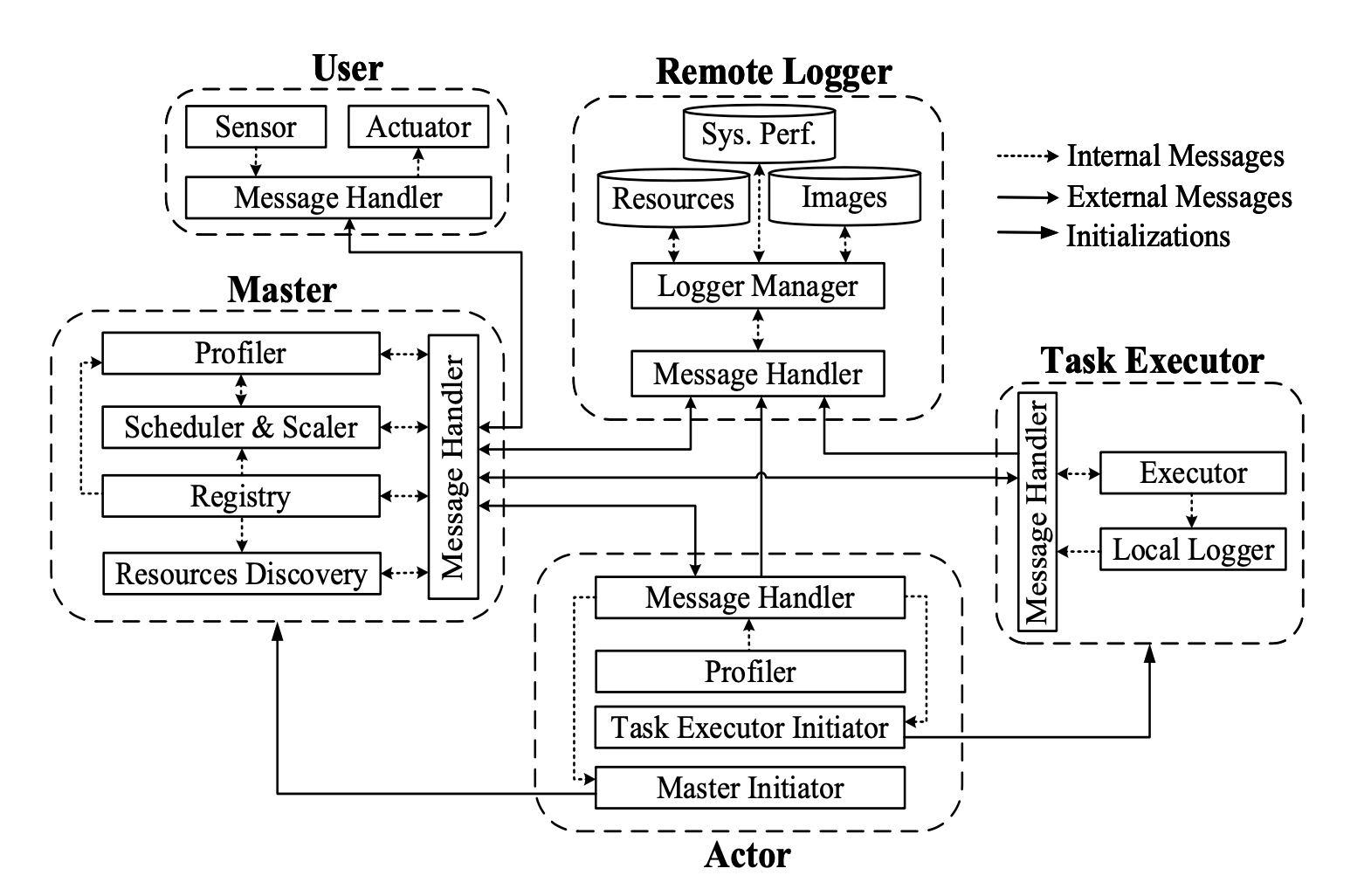}
  \caption{Key components of FogBus2 \cite{deng2021fogbus2}}
  \label{Fig:FB_1}
\end{figure}

FogBus2 \cite{deng2021fogbus2} is a lightweight distributed container-based framework, developed from scratch using Python programming language, enabling distributed resource management in hybrid computing environments. It integrates edge and cloud environments to implement multiple scheduling policies for scheduling heterogeneous IoT applications. In addition, it proposes an optimized genetic algorithm for fast convergence of resource discovery to implement a scalable mechanism that addresses the problem that the number of IoT devices increases or resources become overloaded. Besides, the dynamic resource discovery mechanism of FogBus2 facilitates the rapid addition of new entities to the system. Currently, several resource management policies and IoT applications are integrated with this framework. FogBus2 contains five key containerized components, namely \textit{Master}, \textit{Actor}, \textit{RemoteLogger}, \textit{TaskExecutor}, and \textit{User}, which are briefly described below.
\begin{itemize}
\item \textit{\textbf{Master}}: It handles resource management mechanisms such as scheduling, scalability, resource discovery, registry, and profiling. It also manages the execution of requested IoT applications. 
\item \textit{\textbf{Actor}}: It manages the physical resources of the node on which it is running. Also, it receives commands from the \textit{Master} component and initiates the appropriate \textit{Task Executor} components based on each requested IoT application. 
\item \textit{\textbf{Remote Logger}}: It collects periodic or event-driven logs from other components (e.g., profiling logs, performance metrics) and stores them in persistent storage using either a file system or database.
\item \textit{\textbf{TaskExecutor}}: Each IoT application consists of several dependent or independent tasks. The logic of each task is containerized in one \textit{TaskExecutor}. Accordingly, it executes the corresponding task of the application and can be efficiently reused for other requests of the same type.
\item \textit{\textbf{User}}: It runs on the user's IoT device and handles the raw data received from sensors and processed data from \textit{Master}. It also sends placement requests to the \textit{Master} component for the initiation of an IoT application. 
\end{itemize}
Figure~\ref{Fig:FB_1} shows an overview of the five main components of the FogBus2 and their interactions.

\subsection{K3s: Lightweight Kubernetes}
\begin{figure}[ht]
  \centering
  \includegraphics[width=\linewidth]{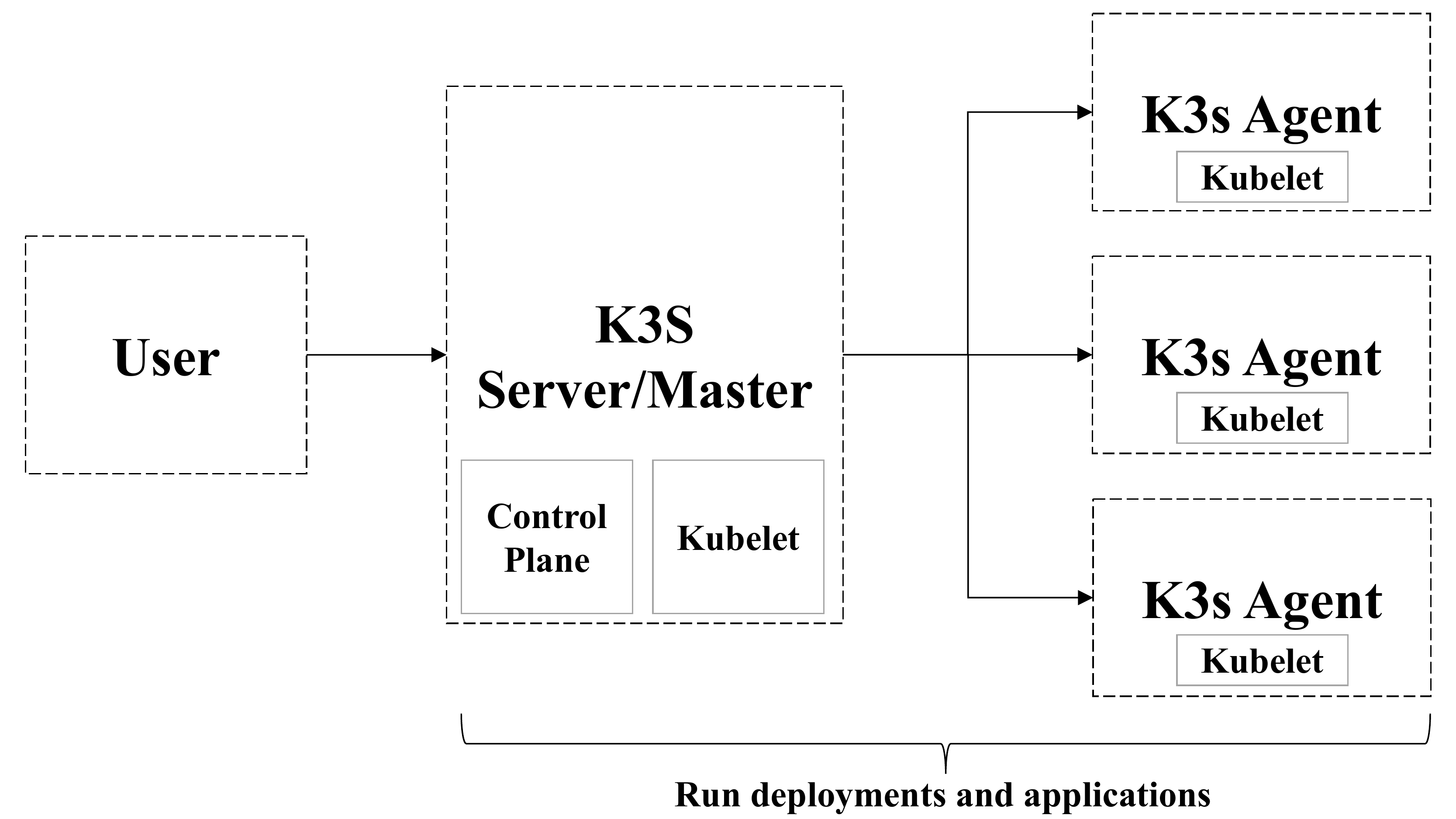}
  \caption{The architecture of a single server K3s cluster}
  \label{Fig:FB_2}
\end{figure}

K3s is a lightweight orchestration tool designed for resource-limited environments, suitable for IoT and Edge/Fog computing \cite{K3SLK}. Compared to Kubernetes, K3s is half the size in terms of memory footprint, but API consistency and functionality are not compromised \cite{todorov2021design}. Figure~\ref{Fig:FB_2} shows the architecture of a K3s cluster containing one server and multiple agents. Users manage the entire system through the K3s server and make appropriate usage of the resources of the K3s agents in the cluster to achieve optimal operation of applications and services.
K3s clusters allow pods (i.e., the smallest deployment unit) to be scheduled and managed on any node. Similar to Kubernetes, K3s clusters also contain two types of nodes, with the server running the control plane components and kubelet (i.e., the agent that runs on each node), and the agent running only the kubelet \cite{K3S}. Typically a K3s cluster carries a server and multiple agents. When the URL of a server is passed to a node, that node becomes an agent; otherwise, it is a server in a separate K3s cluster \cite{K3S,K3SLK}.
\subsection{Related Work}
Rodriguez et al. \cite{rodriguez2019container} investigates multiple container orchestration tools and proposes a taxonomy of different mechanisms that can be used to cope with fault tolerance, availability, scalability, etc. Zhong et al. \cite{zhong2020cost} proposed a Kubernetes-based container orchestration technique for cost-effective container orchestration in cloud environments. The FLEDGE, developed by Goethals et al. \cite{goethals2019fledge}, implements container orchestration in an Edge environment that is compatible with Kubernetes. Pires et al. \cite{pires2021distributed} proposed a framework, named Caravela, that employs a decentralized architecture, resource discovery, and scheduling algorithms. It leverages users' voluntary Edge resources to build an independent environment where applications can be deployed using standard Docker containers. Alam et al. \cite{alam2018orchestration} proposed a modular architecture that runs on heterogeneous nodes. Based on lightweight virtualization, it creates a dynamic system by combining modularity with the orchestration provided by the Docker Swarm. Ermolenko et al. \cite{ermolenko2021internet} studied a framework for deploying IoT applications based on Kubernetes in the Edge-Cloud environment. It achieves lightweight scaling of task-based applications and allows the addition of external data warehouses.
\par
In the current literature, some techniques such as \cite{ermolenko2021internet,zhong2020cost} use Kubernetes directly on Edge/Fog nodes, which have a high overhead on resource-limited Edge/Fog nodes. Some techniques such as \cite{alam2018orchestration} are restricted to run a master node (i.e., server) only on the Cloud, which does not support different cluster deployment approaches. Moreover, some orchestration techniques such as \cite{pires2021distributed} are only working with nodes with public IP addresses, which restricts many use-cases in Edge/Fog computing environments where nodes do not have public IP addresses. Considering the current literature, there exists no lightweight container orchestration technique for the complete deployment of containerized resource management frameworks in hybrid computing environments, where heterogeneous nodes are distributed in Edge/Fog and cloud computing environments. 
\section{Container Orchestration Approach}
\label{Sec:Orchestration}
In this section, we propose a feasible approach for deploying container orchestration techniques in hybrid computing environments. First, we present a high-level overview of the design. Next, we introduce the concrete implementation details of the proposed approach.
\subsection{Overview of the Design}
\begin{figure}[!t]
  \centering
  \includegraphics[width=\linewidth]{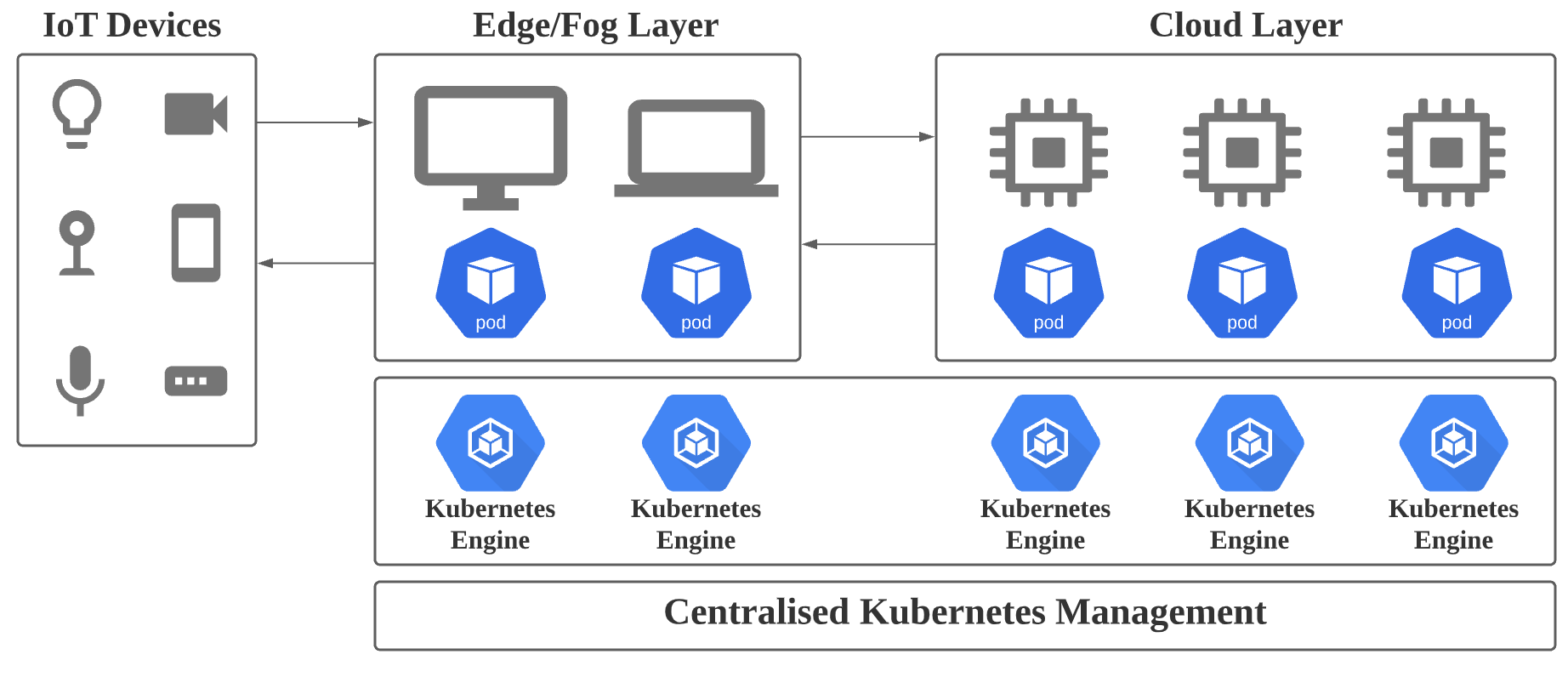}
  \caption{Overview of the design pattern}
  \label{Fig:FB_3}
\end{figure}

To build a complete hybrid computing environment for different IoT scenarios, we use several Cloud and Edge/Fog nodes. We choose K3s as the backbone for the hybrid computing environment because it only occupies less than half of the resources of Kubernetes, but fully implements the Kubernetes API, and is specially optimized for the resource-constrained nodes at the Edge/Fog layer. In practice, we use three Cloud instances at the Cloud layer and create several Linux virtual machines (aligned with hardware specification of Raspberrypi Zero) as our Edge/Fog nodes. Our Cloud nodes have public IP addresses while Edge/Fog nodes do not hold public IP addresses. To address this problem, we use Wireguard to set up a lightweight Peer-to-Peer (P2P) VPN connection among all nodes. After creating the hybrid computing environment, we start to embed the FogBus2 resource management framework into it. To take advantage of the container orchestration tool, we allocate only one container to each Pod created by K3s, with only one component of the FogBus2 framework running inside each container. Also, to balance the load on each node between clusters, we assign pods to different nodes. The initialization of the FogBus2 components requires the binding of the host IP address, which will be used to pass information between the different components. This means that in K3s clustering, the FogBus2 component needs to bind the IP address of the pod, which poses a difficulty for the implementation, as usually the pod is created at the same time as the application is deployed. To address this challenge, we evaluate three approaches and finally decide to use host network mode to deploy the FogBus2 framework in the K3s hybrid environment. Host network mode allows pods to use the network configuration of virtual instances or physical hosts directly, which addresses the communication challenge of the FogBus2 components and the conflict between K3s network planning service and VPN. Figure~\ref{Fig:FB_3} shows a high-level overview of our proposed design pattern.
\begin{table*}[!t]
  \caption{Configuration of Nodes in Integrated Computing Environment}
  \resizebox{\textwidth}{!}{
  \begin{tabular}{c|c|c|c|c|c|c|c}
    \toprule
    Node Tag & Node Name & Computing Layer & Specifications & Public IP & Private IP & Port & Preparation\\
    \midrule
    \texttt A &	Nectar1	& Cloud & 16-core CPU, 64GB RAM & 45.113.235.156 & 192.0.0.1 & auto assign &	docker\\
    \texttt B &	Nectar2	& Cloud & 2-core CPU, 9GB RAM & 45.113.232.199 & 192.0.0.2 & auto assign &	docker\\
    \texttt C &	Nectar3	& Cloud & 2-core CPU, 9GB RAM & 45.113.232.232 & 192.0.0.3 & auto assign &	docker\\
    \texttt D &	VM1 & Edge & 1-core CPU, 512MB RAM & - & 192.0.0.4 & auto assign & docker\\
    \texttt E &	VM2 & Edge & 1-core CPU, 512MB RAM & - & 192.0.0.5 & auto assign & docker\\
    \bottomrule
  \end{tabular}
  }
  \label{Tab:FB_1}
\end{table*}
\subsection{Configuration of Nodes}
The deployed hybrid computing environment consists of several instances, labeled A through E. The node list, computing layer, specifications, public network IP address, and private network IP address, after the VPN connection is established, are shown in Table~\ref{Tab:FB_1}.
\subsection{P2P VPN Establishment}
\begin{figure}[!ht]
  \centering
  \includegraphics[width=0.7\linewidth]{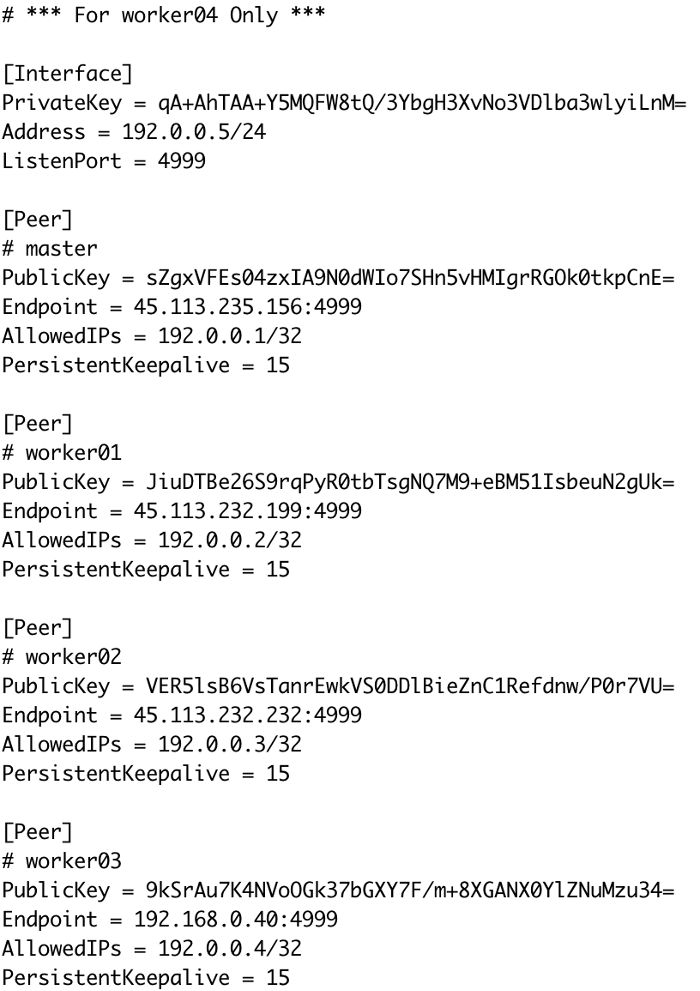}
  \caption{A sample configuration script for the Wireguard}
  \label{Fig:FB_4}
\end{figure}

As shown in Table~\ref{Tab:FB_1}, Cloud nodes have public IP addresses, while in most cases, devices in the Edge/Fog environment do not have public IP addresses. In this case, in order to build a hybrid computing environment, we need to establish a VPN connection to integrate the Cloud and Edge/Fog nodes. We use Wireguard to establish a lightweight P2P VPN connection between all the nodes. In the implementation, we install the Wireguard on each node and generate the corresponding configuration scripts (based on the FogBus2 VPN scripts) to ensure that each node has direct access to all other nodes in the cluster. A sample configuration script for the Wireguard VPN, derived from FogBus2 scripts, is shown in Figure~\ref{Fig:FB_4}.
\subsection{K3s Deployment}
\begin{figure}[!t]
  \centering
  \includegraphics[width=0.7\linewidth]{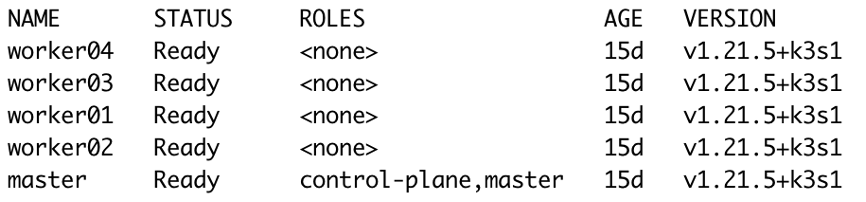}
  \caption{K3s deployment in computing environment}
  \label{Fig:FB_5}
\end{figure}

The K3s server can be located at the Cloud or at the edge, while the remaining four nodes act as K3s agents. As the aim of this research is to enable container orchestration on the FogBus2 framework, we need to install and enable Docker on both the server and agents before building K3s. First, we install and start the K3s server in Docker mode. K3s allows users to choose the appropriate container tool, but as all components of FogBus2 run natively in Docker containers, we use Docker mode to initialize the K3s server to allow it to access the Docker images. Then, we extract a token from the server, which will be used to join other agents to the server. After that, we install the K3s on each agent, specifying the IP of the server and the token obtained from the server during installation to ensure that all agents can properly connect to the server. Figure~\ref{Fig:FB_5} shows the successful deployment of the K3s cluster.
\subsection{Fogbus2 Framework Integration}
In the native design of the FogBus2 framework, all components are running in containers. The pod, as the smallest unit created and deployed by K3s, can wrap one or more containers. Any containers in the same pod will share the same namespace and local network. Containers can easily communicate with other containers in the same or different pod as if they were on the same machine while maintaining a degree of isolation. So first, we are faced with the choice of assigning only one container per pod (i.e., a component that the FogBus2 framework is built on) or allowing each pod to manage multiple containers. The former design would balance the load on K3s nodes as much as possible to facilitate better management by the controller. While the latter design would reduce the time taken to communicate between components and provide faster feedback to users. We decide to adopt the former design to achieve batch orchestration and self-healing from failures.
\par
In order to integrate all types of FogBus2 framework's components into K3s, we first define the YAML deployment files for necessary components. This file is used to provide the object's statute, which describes the expected state of the object, as well as some basic information about the object. In our work, the YAML deployment file serves to declare the number of replicas of the pod, the node it is built on, the name of the image, the image pulling policy, the parameters for application initialization, and the location of the mounted volumes. Code Snippet~\ref{Sc:master} illustrates the YAML deployment file for the \textit{Master} component of the FogBus2 framework.
\renewcommand{\lstlistingname}{Code Snippet}
\begin{lstlisting}[language=Python,caption={The YAML deployment file for the \textit{Master} component of the FogBus2 framework},label=Sc:master]
# YAML deployment file for the Master component 
# of the FogBus2 framework
apiVersion: apps/v1
kind: Deployment
metadata:
  labels:
    app: fogbus2-master
  name: fogbus2-master
spec:
  replicas: 1
  selector:
    matchLabels:
      app: fogbus2-master
  strategy:
    type: Recreate
  template:
    metadata:
      labels:
        app: fogbus2-master
    spec:
      containers:
      - env:
        - name: PGID
          value: "1000"
        - name: PUID
          value: "1000"
        - name: PYTHONUNBUFFERED
          value: "0"
        - name: TZ
          value: Australia/Melbourne
        image: cloudslab/fogbus2-remote_logger
        imagePullPolicy: ""
        name: fogbus2-master
        args: ["--bindIP", "192.0.0.1", 
              "--bindPort", "5001", 
              "--remoteLoggerIP", "192.0.0.1", 
              "--remoteLoggerPort", "5000", 
              "--schedulerName", "RoundRobin", 
              "--containerName", 
              "TempContainerName"]
        resources: {}
        volumeMounts:
        - mountPath: /var/run/docker.sock
          name: fogbus2-master-hostpath0
        - mountPath: /workplace/
          name: fogbus2-master-hostpath1
        - mountPath: /workplace/.mysql.env
          name: fogbus2-master-hostpath2
      restartPolicy: Always
      serviceAccountName: ""
      nodeName: master
      hostNetwork: true
      volumes:
      - hostPath:
          path: /var/run/docker.sock
        name: fogbus2-master-hostpath0
      - hostPath:
          path: /home/hehe/FogBus2/containers
                /master/sources
        name: fogbus2-master-hostpath1
      - hostPath:
          path: /home/hehe/FogBus2/containers
                /master/sources/.mysql.env
        name: fogbus2-master-hostpath2
status: {}
\end{lstlisting}

In the communication design of the FogBus2 framework, the initialization of components requires the binding of the host IP address, which will be used to pass information between components. For example, when a \textit{Master} component is created, the IP address of the host will be passed in as a required parameter, which will also be passed in as a necessary parameter to initializing the \textit{Actor} component. Because the FogBus2 framework has some generic functions that will be used by multiple or all components, the \textit{Master} component will send its assigned host/VPN IP address to the \textit{Actor} component and requests to return the information to this address. If this IP address is not the same as the IP address used to initialize the \textit{Actor} component, communication can not be correctly established. When the FogBus2 framework is deployed using Docker Compose (e.g., the native way that FogBus2 is deployed), communication between the components is smooth because the containers are running directly on the host. However, when the FogBus2 framework starts in K3s, the communication mechanism between the components should be updated since containers are running in pods and each pod has its own IP address. Components cannot listen to the IP address of the host because, by default, the pod's network environment is separate from the host, which poses a challenge for the deployment of the FogBus2 framework. To cope with this problem, we propose the following three design models.
\subsubsection{Host Network}
When starting FogBus2 components in a K3s cluster, instead of using the cluster's own network services, we use the host's network configuration directly. Specifically, we connect each pod directly to the network of its host. In this case, the components in the FogBus2 framework can be bound directly to the host's network at initialization, and the IP address notified to the target component is the same as the one configured by the target component at initialization. This design implements the following functions for the FogBus2 framework:
\begin{itemize}
\item \textbf{Batch Orchestration}: It allows containers to be orchestrated across multiple hosts. In contrast, the native FogBus2 uses docker-compose, which can only create a single container instance locally. 
\item \textbf{Health Check}: The system knows when the container is ready and can start accepting traffic.
\item \textbf{Self-healing from Failure}: When a running pod stops abnormally or is deleted by mistake, the system can restart the pod. 
\item \textbf{Dynamic Change}: Users can dynamically change the resources limit of the running pods, including the size of physical memory footprint, the number of physical CPUs, etc.
\item \textbf{Resource Utilization}: The system can distribute each application on each node, and choose the one with the lowest physical resource usage to deploy.
\end{itemize}
\par
However, this design pattern sacrifices some of the functionality of K3s. When pods are connected directly to the network environment where the hosts are located, the K3s controller will not be able to optimally manage all the containers within the cluster because these services require the K3s controller to have the highest level of access to the network services used by the pods. If the pods are on a VPN network, we will not be able to implement all the features of K3s. We use Host Network mode to deploy the FogBus2 framework in the K3s cluster in this paper. 
\subsubsection{Proxy Server}
As the problem stems from a conflict between the communication design of the FogBus2 framework and the communication model between pods in the K3s cluster, we can create a proxy server that defines the appropriate routing policies to receive and forward messages from different applications. When a FogBus2 component needs to send a message to another component, we import the message into the proxy server, which analyses the message to extract the destination and forward it to the IP address of the target component according to its internal routing policy. This approach bypasses the native communication model of the FogBus2 framework, and all communication between applications is done through the proxy server. 
\par
There are two types of communication methods in the FogBus2 framework, proprietary methods and generic methods. The proprietary methods are used to communicate with fixed components, such as master and remote logger, whose IP addresses are configured and stored as global variables when most components are initialized. In contrast, the generic methods are used by all components and are called by components to transmit their IP addresses as part of the message for the target component. Therefore, to enable all components to send messages to the proxy server for processing, we need to change the source code of the FogBus2 framework so that all components are informed of the IP address of the proxy server at initialization, and to unify the two types of communication methods so that components will include information about the target in the message and send it to the proxy server. As a result, this design would involve a redesign of the communication model of the FogBus2 framework.
\subsubsection{Environment Variable}
In the K3s cluster, when the application is deployed, the cluster controller will automatically create a pod to manage the container in which the application resides. However, in the YAML file, we can obtain the IP address of the created pod when configuring the container information, which allows us to pass it in as an environment variable when initializing the components of the FogBus2 framework. Then, the IP address bound to the component is the IP address of the pod and the component can transmit this address to the target component when communicating and receiving a message back. 
\par
However, in our experiments, we find that pods on different nodes have problems communicating at runtime. We trace the flow of information transmitted and find that the reason for this is the conflict between the network services configured within the cluster and the VPN used to build the hybrid computing environment. The pods possess unique IP addresses and use them to communicate with each other, but these addresses cannot be recognized by the VPN on the nodes, which prevents the information from being transferred from the hosts. To solve this problem, we have proposed two solutions:
\begin{itemize}
\item \textbf{Solution 1}: K3s uses flannel as the Container Network Interface (CNI) by default. We can change the default network service configuration of the K3s cluster and override the default flannel interface with the Wireguard Ethereum Name Service. 
\item \textbf{Solution 2}: We can change the Wireguard settings to add the interface of the network service created by the K3s controller to the VPN profile to allow incoming or outgoing messages from a specific range of IP addresses.
\end{itemize}
%
\section{Performance Evaluation}
\label{Sec:performance_evaluation}
In this section, two experiments are conducted using three real-time applications to evaluate the performance of orchestrated FogBus2 (O-FogBus2) and native FogBus2, as well as the performance of FogBus2 in the hybrid versus cloud environment. The real-time applications used in the experiments are described in Table~\ref{Tab:FB_2}.
\begin{table*}[!t]
  \caption{The list of applications}
  \resizebox{\textwidth}{!}{\begin{tabular}{c|c|c}
    \toprule
    Application Name & Tag & Description\\
    \midrule
    \texttt NaiveFormulaSerialized & Formula & A mathematical formula where different parts are calculated as different tasks.\\
    \texttt FaceDetection (480P Res) & FD480 & Face detection from real-time/recorded video streams at 480P resolution.\\
    \texttt FaceDetection (240P Res) & FD240 & Face detection from real-time/recorded video streams at 240P resolution.\\
    \bottomrule
  \end{tabular}}
  \label{Tab:FB_2}
\end{table*}
\subsection{Experiment 1: Orchestrated FogBus2 vs Native FogBus2}
\label{Section:e1}
This experiment studies the performance of the FogBus2 framework deployed in K3s and compares it with the native FogBus2. In the experiment, we run the systems in the same network environment and set the same scheduling policy to ensure the reliability of the experimental results.
\par
The environment setup for this experiment is shown in Table~\ref{Tab:FB_1}. For both deployment types, we implement the same deployment strategy to ensure fairness, with the \textit{Master} and one \textit{Actor} running on the Edge, and the \textit{Remote Logger} and two other \textit{Actors} running on the Cloud.
\par
\begin{figure*}[!t]
\begin{subfigure}{0.32\textwidth}
  \centering
  \includegraphics[width=\linewidth]{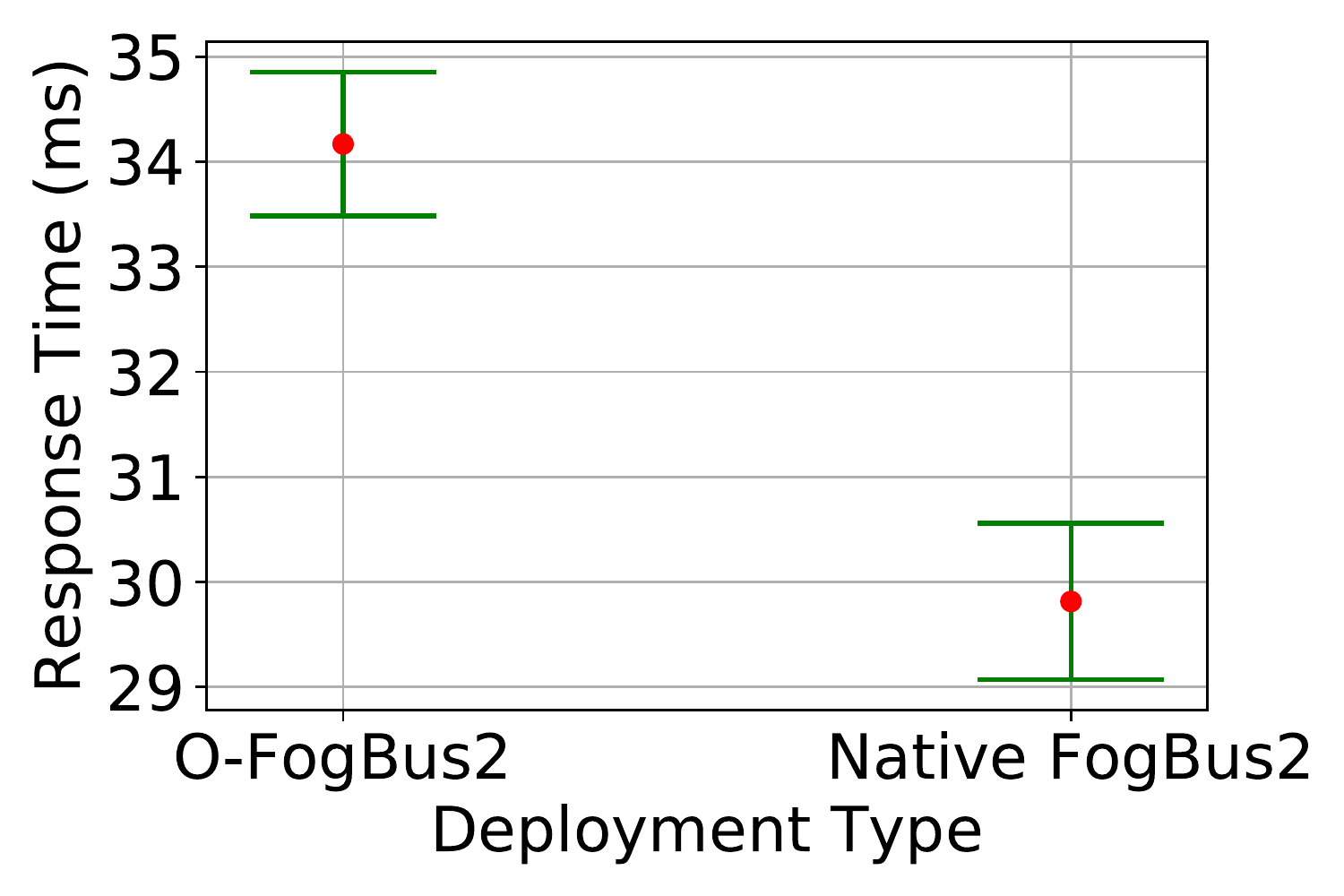}\quad
  \caption{Formula}
  \label{fig:1}
\end{subfigure}%
\begin{subfigure}{0.32\textwidth}
  \centering
  \includegraphics[width=\linewidth]{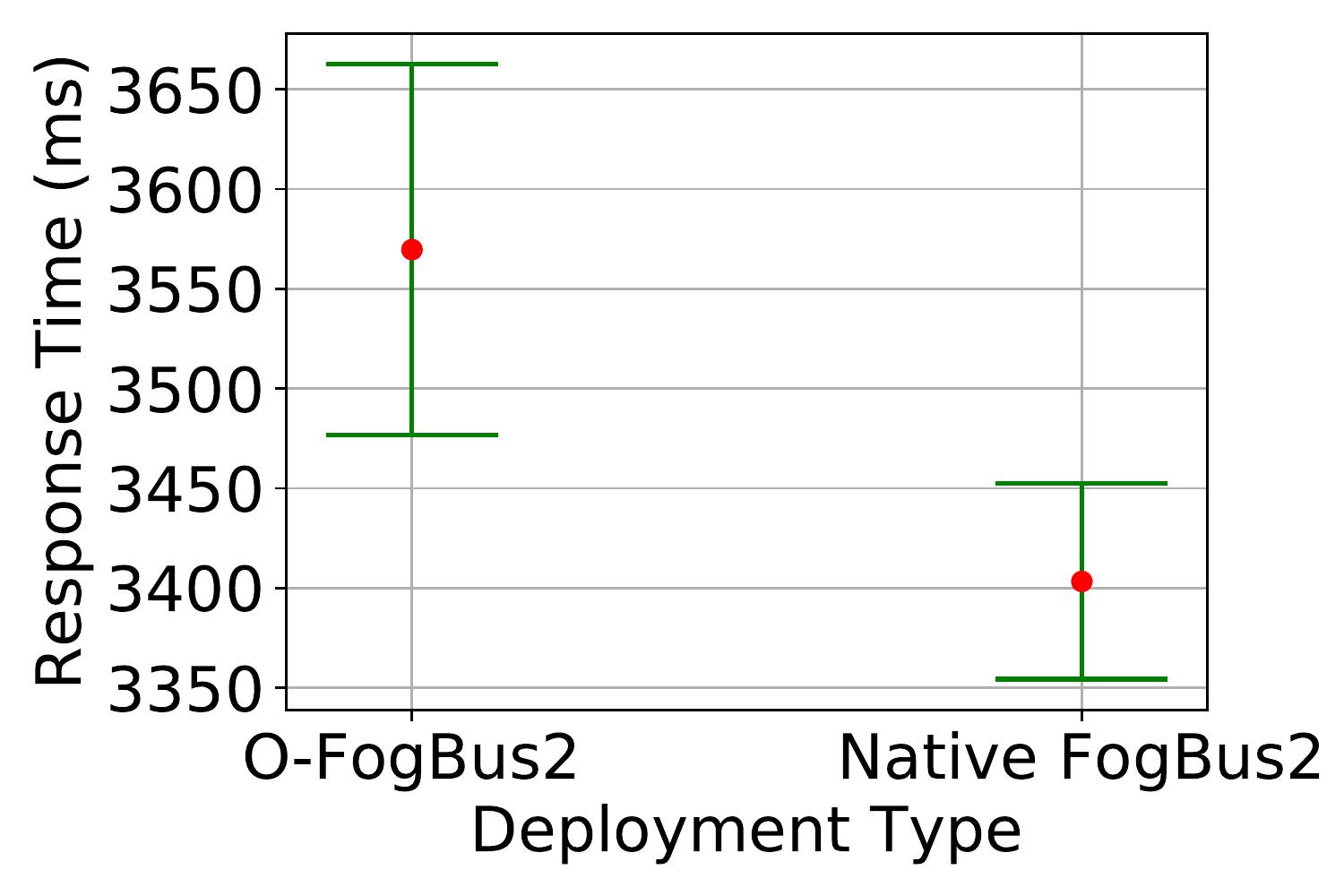}\quad
  \caption{FD480}
  \label{fig:2}
\end{subfigure}
\begin{subfigure}{0.32\textwidth}
  \centering
  \includegraphics[width=\linewidth]{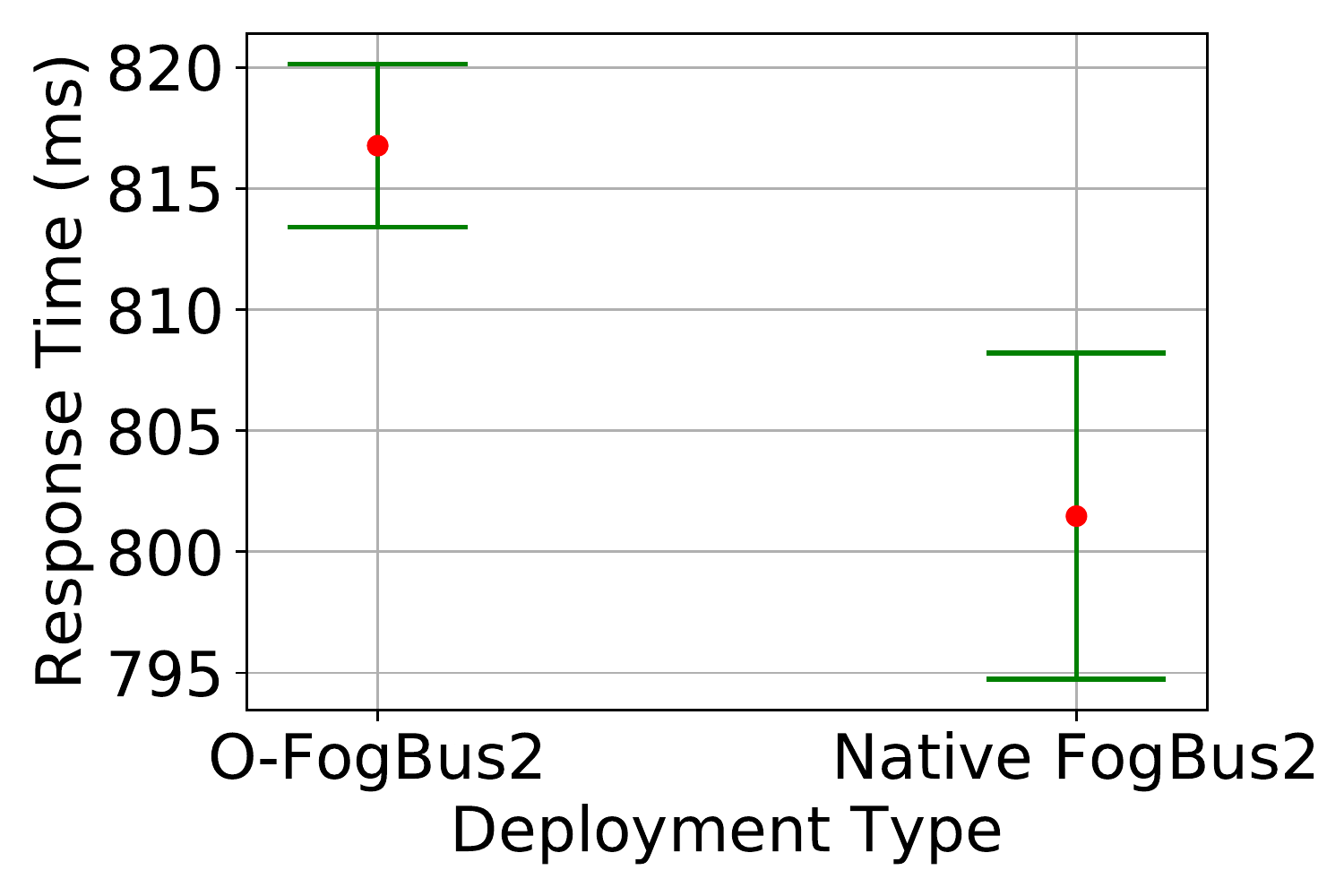}\quad 
  \caption{FD240}
  \label{fig:3}
\end{subfigure}
\caption{Response times for Orchestrated FogBus2 (O-FogBus2) versus native FogBus2 in three applications}
\label{Fig:images1}
\end{figure*}

Figure~\ref{Fig:images1} shows the response time for orchestrated FogBus2 and native FogBus2 using three applications. The red dots represent the average response time, while the top and bottom green lines represent the 95\% confidence interval for the mean value. For all tested applications, when FogBus2 is running in K3s, the average response time is longer than the native FogBus2 framework by an average of 7\%. This is because the management of deployments by the K3s cluster itself requires some overhead, however, given the resource management mechanisms, scheduling, and automatic container health checks provided by K3s, we believe this overhead is very lightweight and acceptable.
\subsection{Experiment 2: Hybrid Environment vs Cloud Environment}
This experiment studies the performance of O-FogBus2 deployed in the hybrid computing environment versus the Cloud computing environment. Same as Section~\ref{Section:e1}, the environment setup for this experiment is shown in Table~\ref{Tab:FB_1}. For the hybrid computing environment, the \textit{Master} and one \textit{Actor} are running on the Edge, and the \textit{Remote Logger} and two other \textit{Actors} are running on the Cloud. And for the Cloud environment, all the components are running on the Cloud.
\par
\begin{figure*}[!t]
\begin{subfigure}{0.32\textwidth}
  \centering
  \includegraphics[width=\linewidth]{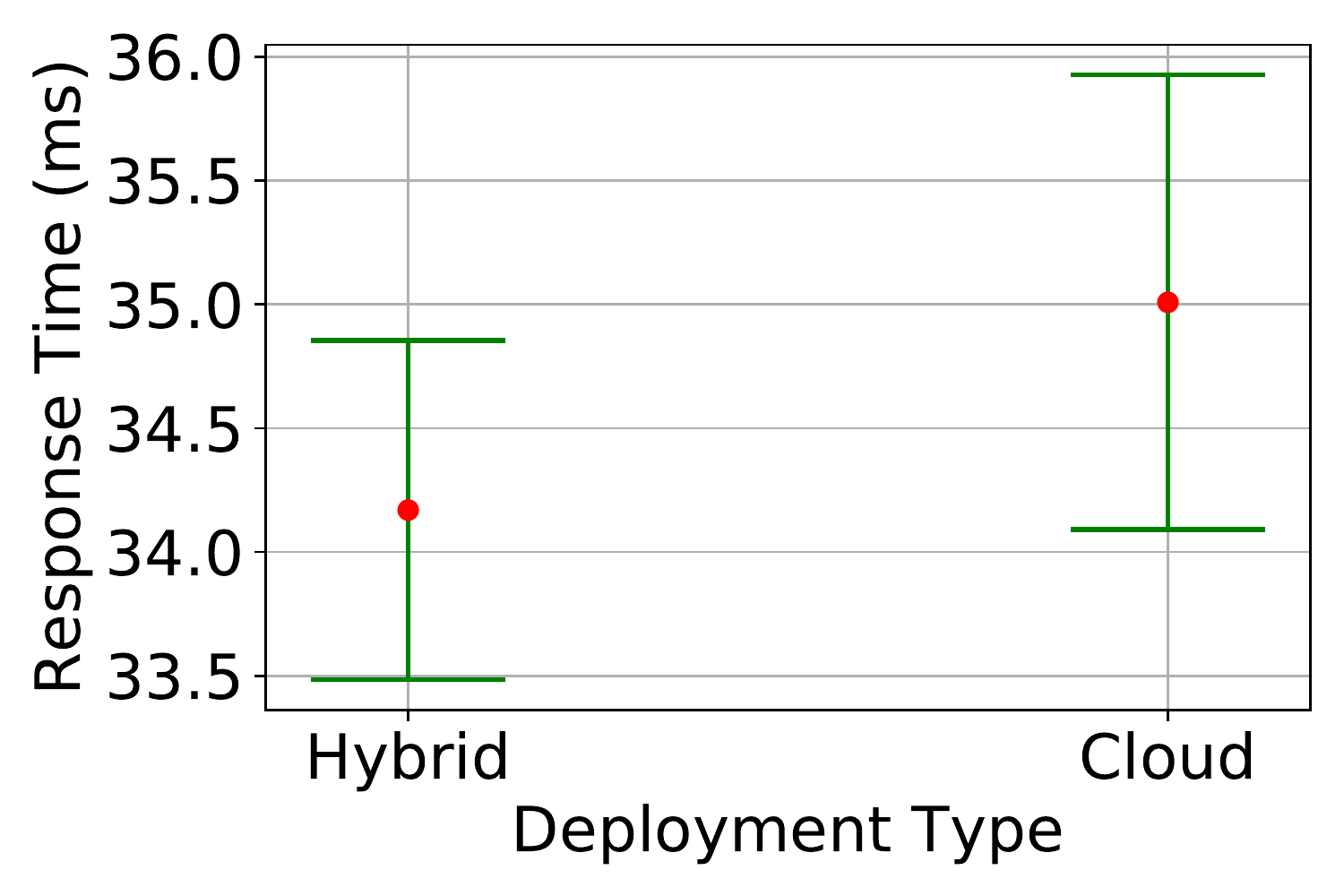}\quad
  \caption{Formula}
  \label{fig:4}
\end{subfigure}%
\begin{subfigure}{0.32\textwidth}
  \centering
  \includegraphics[width=\linewidth]{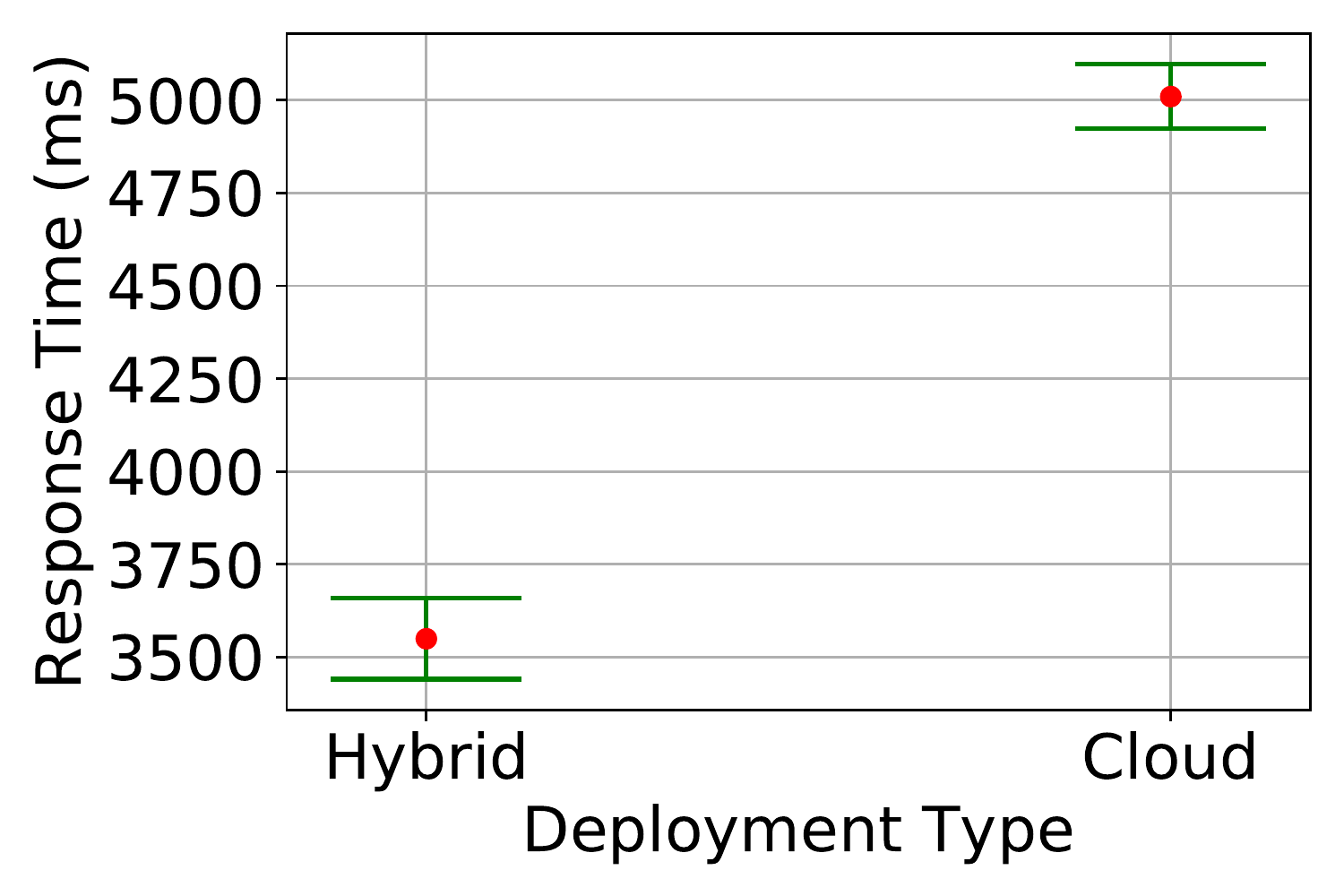}\quad
  \caption{FD480}
  \label{fig:5}
\end{subfigure}
\begin{subfigure}{0.32\textwidth}
  \centering
  \includegraphics[width=\linewidth]{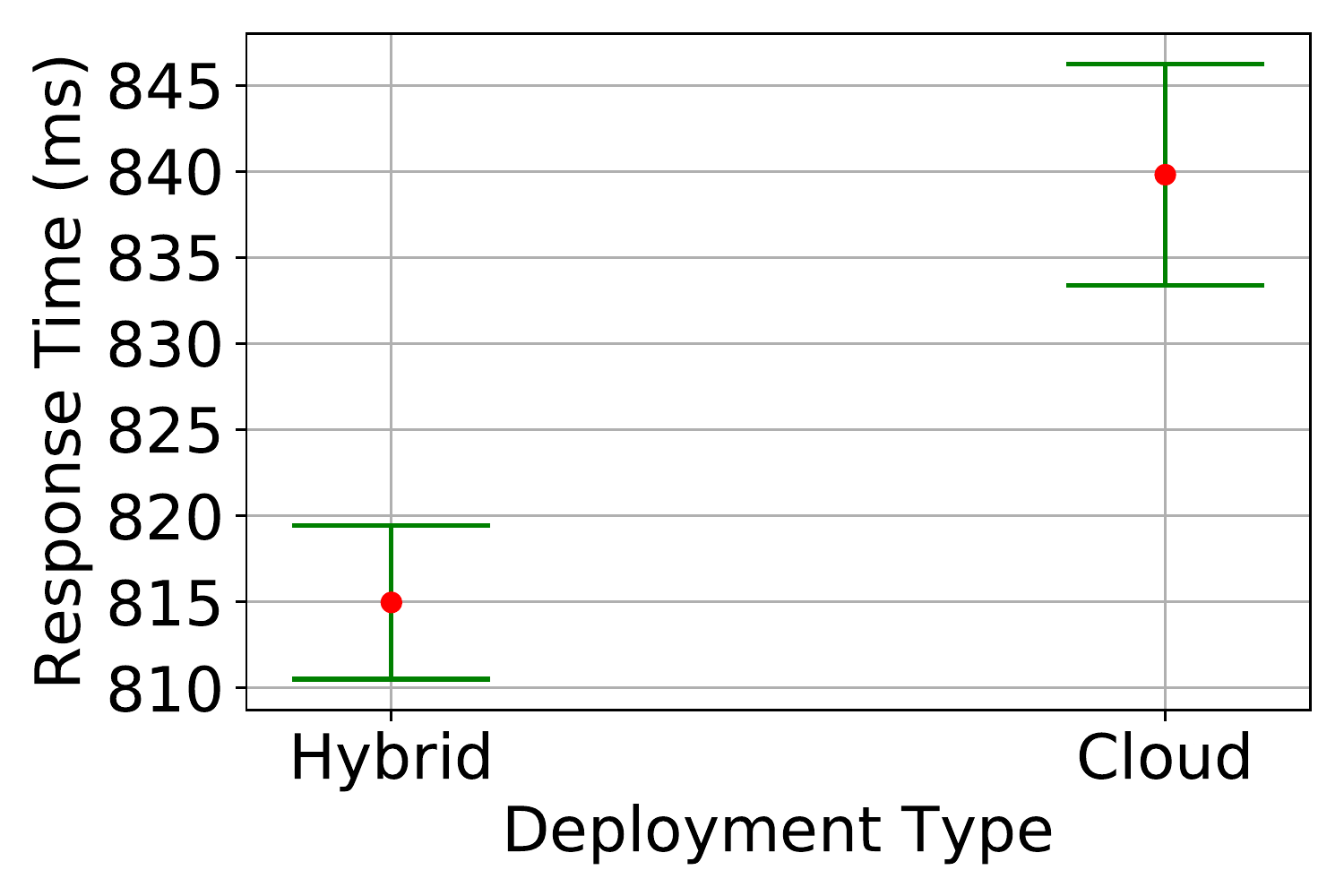}\quad 
  \caption{FD240}
  \label{fig:6}
\end{subfigure}
\caption{Response times for Orchestrated FogBus2 (O-FogBus2) in Hybrid versus Cloud deployment}
\label{Fig:images2}
\end{figure*}

Figure~\ref{Fig:images2} depicts the response time of FogBus2 deployed in hybrid and Cloud environments for three applications. For all tested applications, the average response time is shorter by up to 29\% when FogBus2 is running in the hybrid environment than when FogBus2 is running in the Cloud. This is because the end-users are usually located at the edge of the network and the final result should be forwarded to them. If all the components of FogBus2 are running in the Cloud, it will take longer and will face the impact of the unstable Wide Area Network (WAN). Since FogBus2 is designed for IoT devices to integrate Cloud and Edge/Fog environments, the introduction of K3 does not deprive this function, so we believe that placing the entire system in a hybrid computing environment can reasonably utilize the Cloud and Edge/Fog computing resources and improve system performance.
\section{Conclusions and Future Work}
\label{Sec:conclusion}
In this paper, we discussed the importance of resource management to support real-time IoT applications. We presented feasible designs for implementing container orchestration techniques in hybrid computing environments. This study proposed three design patterns for deploying the containerized resource management frameworks such as the FogBus2 framework into the hybrid environment. Besides, we described the detailed configuration of K3s deployment and the integration of the FogBus2 framework using the host network approach. The Host Network Pattern connects the components of the cluster to the host network environment, using the native communication model of the FogBus2 framework by masking the internal network environment of the cluster while avoiding the network conflict problems related to VPN. Compared to the native Fogbus2 framework, the new system (i.e., O-FogBus2) enables resource limit control, health check, and self-healing from failure to cope with the ever-changing number and functionality of connected IoT devices.
%
%
\par
We identified several future works to further improve the container orchestration for efficient resource management in hybrid computing environments. Firstly, we can consider implementing elastic scalability to automatically add or remove computing resources according to the demands of IoT applications. To address this challenge, the Proxy Server and Environment Variable design approaches can be investigated to enable dynamic scalability. Secondly, lightweight security mechanisms can be embedded into the container orchestration mechanisms. As IoT devices are highly exposed to users, security and privacy become important. However, the limited resources of Edge/Fog devices create difficulties for the implementation of security mechanisms. Therefore, lightweight security mechanisms to ensure end-to-end integrity and confidentiality of user information can be further investigated. Next, integrating different orchestration tools, including KubeEdge, Docker Swarm, and MicroK8s can be considered as an important future direction. Different orchestration tools may be suitable for different computing environments, so it is essential to find the best application scenarios for them. We can explore the impact of different integrated container orchestration tools for handling real-time and non-real-time IoT applications. Also, a variety of scheduling policies can be implemented to automate application deployment and improve resource usage efficiency for clusters, ranging from heuristics to reinforcement learning techniques \cite{goudarzi2021distributed}. For example, scheduling pods to nodes with smaller memory and CPU footprints to automatically load-balancing on the cluster, or spreading replicative pods across different nodes to avoid severe system failures. Furthermore, since machine learning techniques \cite{goudarzi2021distributed,agarwal2021reinforcement} are becoming mature and widely used in various fields, we can consider integrating them into the Edge/Fog and Cloud computing environment. Machine learning techniques can be used to analyze the state of the current computing environment, improve the system's ability to manage resources and distribute workloads. As current machine learning tools are often designed for powerful servers, future research can optimize them to run on resource-constrained Edge/Fog devices. Finally, the adopted techniques can consider the requirements of specific application domains such as natural disaster management, which significantly affect human life.
%
\bibliographystyle{unsrt}
\bibliography{samplepaper.bbl}

\end{document}